\documentclass[numreferences]{kluwer}    
\usepackage{graphicx}

\begin{document}                                                                                   
\begin{article}

\begin{opening}         
\title{Lagrangian Vorticity and Velocity Measurements in Turbulent Jets} 
\author{Poulain \surname{Cedric}} 
\author{Mazellier \surname{Nicolas}} 
\author{Gervais \surname{Philippe}}
\author{Gagne \surname{Yves}\\}
\author{Baudet \surname{Christophe}\email{baudet@hmg.inpg.fr}}  
\runningauthor{C. POULAIN ET AL.}
\runningtitle{Lagrangian Acoustic Measurements}
\institute{Laboratoire des \'Ecoulements G\'eophysiques et Industriels,
INPG/UJF/CNRS-UMR 5519, 1025, Rue de la Piscine 38041 Grenoble}
\date{}

\begin{abstract}
In this paper we report an experimental investigation of various statistical properties of the spatial Fourier modes of
the vorticity field in turbulent jets for a large range of Reynolds numbers ($530 \leq R_{\lambda} \leq 6100$). 
The continuous time evolution of a spatial Fourier mode of the vorticity distribution, characterized by a well defined wavevector, 
is obtained from acoustic scattering measurements. 
The spatial enstrophy  spectrum, as a function of the spatial wave-vector, is determined by scanning the incoming sound frequencies. 
Time-frequency analysis of the turbulent vorticity fluctuations is also performed for different length scales of the flows. 
Vorticity time-correlations show that the characteristic time of a Fourier mode behaves as the sweeping time. 
Finally, we report preliminary Lagrangian velocity measurements obtained using acoustic scattering by soap bubbles inflated with helium. 
Gathering a very large number of passages of isolated bubbles in the scattering volume, one is able to compute the Lagrangian velocity PDF and 
velocity spectrum. Despite the spatial filtering due to the finite size of the bubble, the latter exhibits a power law with the -2 exponent 
predicted by the Komogorov theory, over one decade of frequencies.
\end{abstract}


\end{opening}           


\section{Introduction}

In fully developped turbulence, most of the experimental, numerical and theoretical works \cite{Frisch95} rely on the statistics of the longitudinal 
velocity increments to study the dynamic of a given  turbulent length scales in the physical space.  In the Fourier space, there are numerous 
theoretical and some numerical studies (\cite{Monin&Yaglom}), but very few experiments. The objective of this paper is mainly to give experimental 
vorticity data in the Fourier space. Firstly, the ultra-sound acoustic scattering method is described, and the experimental conditions 
are detailed. Then, both time and spatial spectra	 of vorticity are presented and are extended to a time-frequency analysis. Correlation 
time of the vorticity Fourier mode is discussed. Finally, we show how the same scattering method can be applied to detect and follow the 
trajectory of isolated soap bubbles in a turbulent air jet, in order to extract Lagrangian velocity data and compute statistics.

\subsection{Acoustic scattering by vorticity}  
Wave propagation (light and sound) in fluids is known to be strongly affected by turbulent velocity and/or temperature gradients. 
In the case of direct propagation, turbulence is responsible for random wave amplitude  and phase fluctuations along the ray paths leading to the well 
known stellar scintillation phenomena limiting the spatial resolution of optical and radio observations \cite{Monin&Yaglom,Goodman}. 
Acoustic wave propagation is also sensitive to local velocity and temperature fluctuations induced by turbulence: such fluctuations are responsible for local 
fluctuations of the sound velocity along the ray paths which result in distorsions of the wave fronts \cite{Monin&Yaglom,Pierce,Morse,Engler89}.\\ 
In the presence of spatial inhomegeneities with length scales comparable to the acoustic wave-length $\lambda_{s}$, an incident acoustic plane wave can also be
scattered giving rise to scattered acoustic waves propagating in directions away from the incident wave direction of propagation \cite{Kov76,Korm80}. 
There exists a large number of theoretical \cite{Obuk53,Kraic53,Chu58,Batch57,Lund89} and numerical \cite{Llewellyn1,Llewellyn2,Colonius94} studies dealing with the acoustic 
scattering phenomenon by velocity fields. Most of it was initiated nearly fifty years ago and started  with the papers  authored by A.M.~Obukhov \cite{Obuk53}, 
R.H.~Kraichnan \cite{Kraic53}, B.~T.ÊChu and L.~S.~G.~Kov\`asznay \cite{Chu58},G.~K.~Batchelor \cite{Batch57} to mention but a few.  
Recently, F. Lund \cite{Lund89} has established, under reasonable and fairly restrictive assumptions, a linear relation between the scattered amplitude 
of a plane  acoustic wave incident on a turbulent flow and the spatial Fourier transform of the vorticity field.\\ 
In analogy with the more usual light scattering phenomenon, and following Batchelor \cite{Batch57} the physical mechanism at the origin of acoustic scattering 
by vorticity can be thought  of as follows: an acoustic wave  impinging on a vorticity distribution induces fluctuations of the  vorticity at the incoming  sound 
frequency (by virtue of the Kelvin circulation theorem).  
Each scatterer (vortex)  acting as a secondary source will, in turn, radiate a sound wave. The coherent average (taking into account the relativepositions  of 
the individual vorticity elements) over the scatterers distribution results in the emission, outside the vorticity domain, of a scattered acoustic waves.  
Note that whereas the light scattering process is usually linear, the acoustic scattering phenomenon depicted here stems from the non-linear term of the 
Navier-Stokes equation, and requires more detailed explanations and computations, beyond the scope of this paper. In particular, one can find in 
\cite{Lund89} a clear explanation of the respective contributions to the total acoustic scattering amplitude of the vorticity field on one hand, and of the 
irrotational velocity field, induced by the vorticity field, on the other. Using a Born approximation, Lund {\it et al.} obtain the following linear relation 
between the scattered acoustic pressure amplitude and the spatial Fourier transform of the vorticity field:

\begin{equation}
\frac{p_{scat}(\nu)}{p_{inc}} = \pi^{2} i \frac{-cos(\theta_{scat})}{1-cos(\theta_{scat})}
\frac{\nu e^{i 2\pi \nu D/c}}{c^{2}D} ({\bf n} \land {\bf r}).
{\bf \Omega}({\bf q}_{scat}, \nu - \nu_{o})
\label{eqn_Diff}
\end{equation}

where $\land$ and ``.'' stand for the vector product and the scalar product respectively, and where the
scattering wave-vector (also called momentum transfert) is given by:

\begin{equation}
{\bf q}_{scat} = \frac{2\pi}{c}(\nu {\bf r} - \nu_o {\bf n}) \simeq 
\frac{4\pi\nu_o}{c}\sin(\frac{\theta_{scat}}{2}) \frac{{\bf r}-{\bf n}}{|{\bf r}-{\bf n}|}
\mbox{ for } \nu \simeq \nu_o
\label{q_scat}
\end{equation}

\begin{figure}[h] 
\vspace{6pc}
\centerline{
\includegraphics[width=10cm]{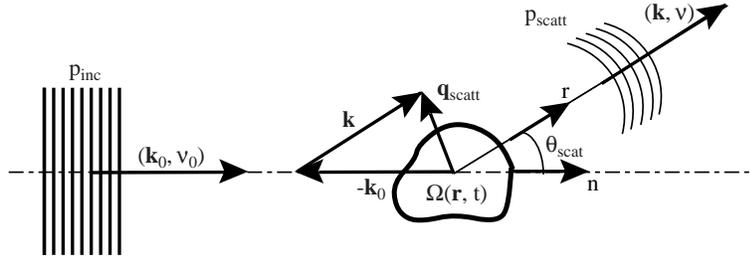}
\caption[]{Acoustic Scattering by a Vorticity Distribution}
}
\label{SchematicScatt}
\end{figure}

In the above equations, the meaning of the different variables is the following according to
the schematic drawing in fig. (1):

\begin{itemize}
\item ${\bf q}_{scat} = {\bf k}-{\bf k}_o$ is  the scattering wave-vector.
\item ${\bf n}$ and ${\bf r}$ are the unit vectors in the incident and  scattered wave directions of propagation,
respectively.
\item ${\bf k}_o$ (resp. $\nu_o$) is the vector wave-number (resp. frequency) of the incoming sound wave (in the
direction ${\bf n}$).
\item ${\bf k}$ (resp. $\nu$) is the vector wave-number (resp. the frequency) of the scattered sound wave (in the
direction of observation ${\bf r}$).
\item $\theta_{scat}$ is the scattering angle.
\item $p_{inc}$ is the complex pressure amplitude of the incoming sound wave (assumed to be plane and
monochromatic).
\item $p_{scat}$  is the complex pressure amplitude of the scattered sound wave.
\item $D$, $c$ and $\lambda_s$ stand respectively for the acoustical path between the measurement area and the detector, the
adiabatic sound velocity and the acoustic wavelength.
\end{itemize}

\subsection{Acoustic scattering: a probe of vorticity} 
 
Equation (\ref{eqn_Diff}), indicates that, using an incident acoustic plane wave (ideally of infinite
extension), one can directly probe the time-space Fourier mode $\Omega_{\perp}({\bf q}_{scat},f=\nu-\nu_{o})$ 
of a well defined component of the vorticity field $\Omega_{\perp}({\bf r},t)$:

\begin{equation}
\Omega_{\perp}({\bf q}_{scat},f) = \int \int \int \int
\Omega_{\perp}({\bf r},t) e^{-j({\bf q}_{scat}{\bf r}-2 \pi f t)} dtd^3r
\label{TFOmega}
\end{equation}

Actually, the direction of the probed vorticity component (indicated by the subscript $_{\perp}$) is
perpendicular to the scattering  plane defined by the vector wave-numbers of the incident and scattered
(detected) acoustic waves. Equation (\ref{eqn_Diff}) can be reformulated as a relation between the
scattered amplitude and the convolution product of the incident acoustic amplitude and 
$\Omega_{\perp}({\bf q}_{scat},f=\nu-\nu_{o})$. Using an inverse Fourier transform on the frequency variable,
one then obtains the following linear relation between the time variables : 

\begin{equation}
p_{scat}({\bf k},t) \propto  L(\theta_{scat})\Omega_{\perp}({\bf q}_{scat}={\bf k}-{\bf k}_o,
t).p_{inc}({\bf k}_o,t)
\label{eqn_Diff_Modul}
\end{equation}

expressing the scattered pressure amplitude as the result of a modulation of the incident acoustic pressure
by the space Fourier transform $\Omega({\bf q}_{scat},t)$. The variations of the angular prefactor
$L(\theta_{scat})$ with the scattering angle $\theta_{scat}$, represented on figure (2), are
typical of a quadrupolar like radiation pattern, diverging at null scattering angle (where the Born approximation breaks down), 
and exhibiting null values at $\theta_{scat}=90 ^{\circ}$ and $\theta_{scat}=180 ^{\circ}$ (back-scattering). 

\begin{figure}[h] 
\vspace{6pc}
\centerline{
\includegraphics[width=12cm]{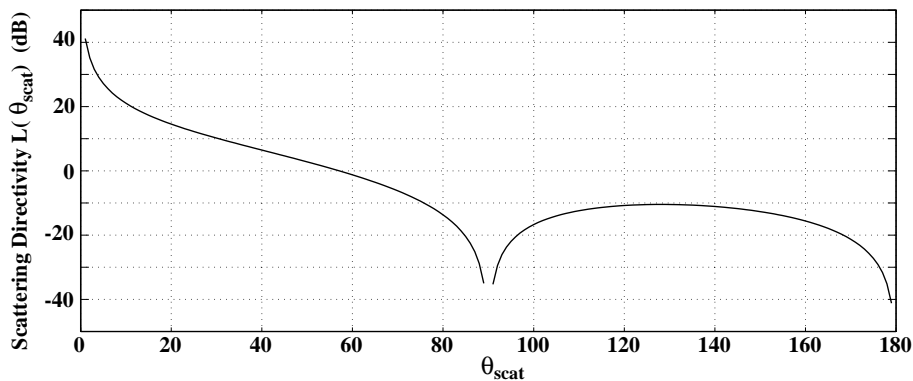}
\caption[]{Acoustic Scattering by Vorticity Directivity}
}
\label{Directiv_Lund}
\end{figure}

Among the several hypothesis put forward in \cite{Chu58,Lund89} to arrive at Eqn. (\ref{eqn_Diff_Modul}), necessary
for the Born approximation to be applicable, the velocity amplitude $u_{sound} = \frac{p_{inc}}{\rho_o c}$ of the incident wave and the Mach
number $M=\frac{v_{flow}}{c}$ of the flow under investigation (associated to the probed vorticity distribution)
must be such that : $u_{sound} \ll v_{flow} \ll c$. In addition, the time scales of the flow are required to be much
smaller than the period $T_o=\frac{1}{\nu_o}$ of the incident wave. These conditions can be easily fulfilled using
ultrasonic acoustic waves with low enough intensity. From a practical point of view, equations~(\ref{eqn_Diff}) and
(\ref{eqn_Diff_Modul}) imply that the scattered pressure signal is a narrow-band signal centered around the
incident frequency $\nu_o$; the continuous time evolution of $\Omega_{\perp}({\bf q}_{scat},t)$ can then be
easily recovered by a simple demodulation operation using  for exemple an heterodyne detection (either analog or
digital) \cite{Papoulis}. After demodulation, for a fixed incoming sound frequency $\nu_o$ and a fixed
scattering angle $\theta_{scat}$ one thus gets the time dynamics of a spatial Fourier mode of the vorticity
$\Omega_{\perp}({\bf q}_{scat},t)$, caracterized by a well defined vector wave-number ${\bf q}_{scat}$
(with specified modulus and orientation).\\ 
The overall effect of the acoustic scattering process can then be viewed as a time-continuous spatial band-pass 
filtering operation of the vorticity distribution $\Omega_{\perp}({\bf r},t)$. 
The center frequency of the spatial filtering operation is given by the scattering 
vector wave-number ${\bf q}_{scat}$ which depends on a combination of the incident frequency $\nu_o$ and of the
scattering angle $\theta_{scat}$ as stated by Eqn~(\ref{q_scat}). For practical reasons and as we dispose of
wide-band transducers (emission and reception), the analysed length-scale of the flow is tuned through the selection 
of an incident frequency $\nu_o$ while the scattering angle is usually kept constant. It is worth noting here that, 
as a result of the demodulation process according to which the instantaneous scattered pressure signal is multiplied by 
both the in phase ($cos(2 \piÊ\nu_o t)$) and  the in quadrature ($sin(2\piÊ\nu_o t)$) signals phase-locked with 
the electric signal driving the sound transmitter we get finally a low frequency signal which is complex (modulus and phase) 
and directly proportionnal to the complex quantity $\Omega_{\perp}({\bf
q}_{scat},t)$, where: 

\begin{equation}
\Omega_{\perp}({\bf q}_{scat},t) = \int \int \int 
\Omega_{\perp}({\bf r},t) e^{-j{\bf q}_{scat}{\bf r}}d^3r
\label{DefOmega_q_t}
\end{equation}

Thus, the instantaneous phase of the demodulated scattered pressure signal is simply a measure of the phase
modulation of the scattered pressure wave with respect to the incident wave (chosen as a phase reference).
Similarly, the instantaneous amplitude of the demodulated scattered pressure signal is a measure of the
amplitude modulation with respect to the constant amplitude of the incident wave. Both informations (phase and
amplitude) are easily recovered from the electric signals owing to the linearity property of the acoustic
transducers used to generate and detect acoustic waves.

\subsection{Spatial filtering: finite size effects and spectral resolution}  

As indicated by equation (\ref{eqn_Diff}) and figure (2), the acoustic amplitude scattered by
vorticity distributions is a strongly decreasing function of the scattering angle $\theta_{scat}$. In order to
preserve an acceptable signal-to-noise ratio, one needs to work with low scattering angles (typ. between $10 ^o$ and
$60 ^o$). Hence, we work with a bistatic configuration, using a pair of acoustic transducers :
one transmitter and one receiver. The relative positions of the two tranducers, as well as their arrangement with
respect to the flow under investigation define the scattering angle and the orientation of the
scattering wave-vector ${\bf q}_{scat}$ with respect to the flow geometry (e.g. its mean velocity). A typical
experimental configuration, for the study of a turbulent jet flow is represented on figure(3). 
Equation (\ref{eqn_Diff}) is valid for an ideal incident plane wave and a perfectly directive detection. 
In a real experiment, transducers are of finite dimensions and one has to take into accountdiffraction effects. 
Let $L$ be the typical size of the acoustic transducers and $\lambda_{s}=\frac{c}{\nu_o}$ the acoustic wavelength, 
at large distances from the transducer (in the far-field limit $D \ge \frac{L^2}{\lambda_{s}}$
\cite{Kinsler82,Pierce}), one expects acoustic energy to be either emitted by the transmitter or detected by the
receiver in a cone corresponding to a continuous distribution of angles defining the angular aperture (or
directivity). In particular, the principal diffraction lobe, around the normal to each tranducer, is a cone with a
half angle aperture $\Delta\theta$ given by \cite{Kinsler82,Pierce}): $\Delta\theta \simeq sin(\Delta\theta) =
\frac{\lambda_{s}}{L}$. By differentiating Eqn. (\ref {q_scat}), one can associate to these distributions of emission
and detection angles, a distribution of probed scattering wave-vectors
${\bf q}_{scat}=4\pi\frac{sin(\frac{\theta_{scat})}{2}}{\lambda_{s}}$ with a typical width: 

\begin{equation}
\Delta q_{scat} = 2\pi\frac{cos(\frac{\theta}{2})}{\lambda_{s}}\Delta\theta \simeq \frac{2\pi}{\frac{L}{cos(\frac{\theta)}{2}}}
\label{Delta_q_scat}
\end{equation}

Equation (\ref {Delta_q_scat}) gives an estimation of the width of spatial band-pass filtering operation performed
by the scattering process around the spatial frequency (length scale) ${\bf q}_{scat}$. It is interesting at this
point to notice that Equation (\ref {Delta_q_scat}) can be interpreted as a mere consequence of the uncertainty
relation $\Delta q_{scat}\Delta x \simeq 2\pi$, where $\Delta x = \frac{L}{cos(\theta)}$ is roughly the extension of
the measurement volume (in the direction of ${\bf q}_{scat}$), defined by the intersection of the incident and
detected (antenna beam) acoustic beams (cf fig. (3)). An equivalent formulation, albeit in
the physical space, of the previous result can be found for example in Ref. (\cite{Pierce} pp. 443--446), where the
scattering amplitude in an experimental bistatic configuration is expressed as: 

\begin{equation}
\Omega_{\perp}({\bf q}_{scat},t) = \int \int \int _{-\infty}^{+\infty}
|F_{rec}({\bf r})F_{trans}({\bf r})|\Omega_{\perp}({\bf r},t) e^{-j{\bf q}_{scat}{\bf r}}d^3r
\label{DefOmega_q_t_FiniteSize}
\end{equation}

where $G({\bf r}) = |F_{rec}({\bf r})(F_{trans}({\bf r})|$ is a ponderation function defined as the modulus of the
cross product of the complex amplitudes of the incident (resp. detected) $F_{rec}({\bf r})$ (resp.
$F_{trans}({\bf r}))$ acoustic beams. 

\begin{figure}[h] 
\vspace{6pc}
\centerline{
\includegraphics[width=12cm]{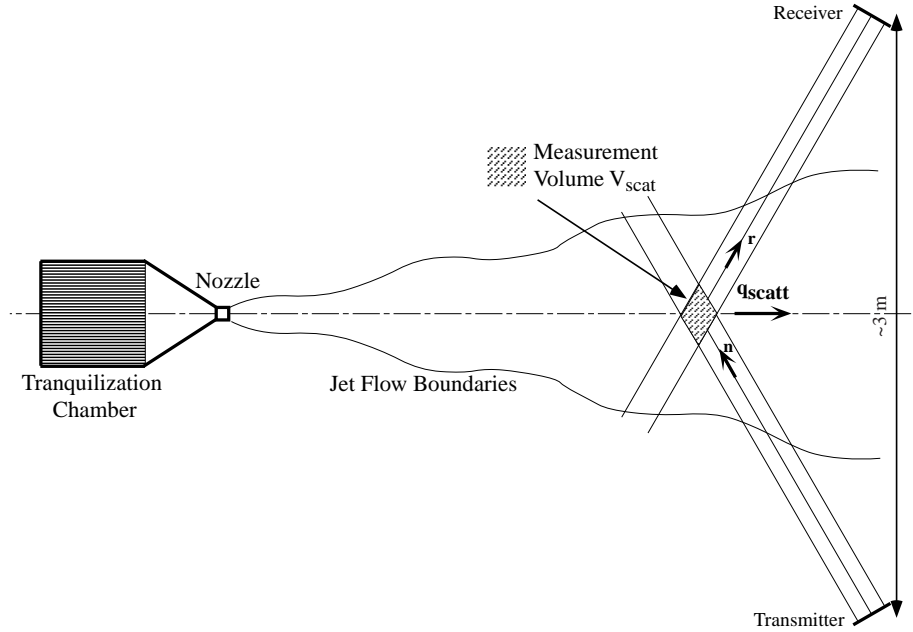}
\caption[]{Acoustic scattering experiment in a turbulent jet flow}
}
\label{Dispositif_Experiment}
\end{figure}

Let's note $V_{scat}$, the volume of the region of space over which
$G({\bf r})$ takes appreciable non zero values and let's call $V_{scat}$ the scattering volume, equation (\ref
{DefOmega_q_t_FiniteSize}) can be expressed as a windowed spatial Fourier transform: 

\begin{equation}
\Omega_{\perp}({\bf q}_{scat},t) = \int\int\int_{V_{scat}} 
\Omega_{\perp}({\bf r},t) e^{-j{\bf q}_{scat}{\bf r}} d^3r
\label{DefOmega_q_t_FiniteSize2DefOmega_q_t_FiniteSize}
\end{equation}


In conclusion, we deal with a complex signal $z_{scat}(t) = \rho(t)e^{j\phi(t)} \propto \Omega_{\perp}({\bf q}_{scat},t)$ with a spectral 
resolution given by : 
\begin{eqnarray}
{\bf q}_{scat} = q_{scat}{\bf e}_x \pm \Delta q_x {\bf e}_x \pm \Delta q_y
{\bf e}_y \pm \Delta q_z {\bf e}_z\\
q_{scat}=\frac{4\pi\nu_o}{c}sin(\theta_{scat}/2)\\
\Delta q_x \simeq \Delta q_y \simeq \Delta q_z \simeq (\frac{1}{V_{scat}})^{1/3} \simeq \frac{1}{\frac{L}{\cos(\theta_{scat}/2)}}
\label{q_scat_SpectralResolution}
\end{eqnarray}

In the following, we shall refer to $p_{scat}(t)$ as the scattered pressure signal (real) with frequencies around $\nu_o$ (high frequency) and to 
$z_{scat}(t)$ as the demodulated scattered pressure signal (complex) with low frequencies.


\section{Experimental Conditions}

\subsection{Flow configurations}

Two axisymetric turbulent jets have been investigated: a laboratory air jet at moderate Reynolds number, (hereafter, called LEGI jet),  
and a low temperature gaseous helium jet at very high Reynolds number which  has been performed in the cryogenic facility hosted by the 
CERN in Geneva. In the laboratory air jet the center of the measurement volume was located on the axis at about 40 diameters downstream 
the nozzle (with a diameter of $12~cm$). The mean velocity and the rms velocity  have been measured in the center of scattering measurement 
volume, with conventional instrumentation (hot wire and Pitot anemometry). The value of the Taylor microscale Reynolds number estimated from 
longitudinal velocity samples and isotropic and homogeneous relations, ranges $530 \le R_{\lambda} \le 785$ \cite{Malecot}.\\
The high Reynolds jet investigated with acoustic scattering \cite{GReC_Adv_Cryo_2002, GReC_PhysicaC_2003} consists in an axisymetric gaseous helium 
jet at a temperature of about $4 ^{\circ} K$. The flow emerges from a $2.5~cm$  diameter nozzle into a large thermally insulated cylindrical chamber 
($4.6~m$ high and with a diameter of $1.4~m$). The  cryostat is connected to a large refrigerator, with $6~kW$ cooling power, allowing flow rates of 
cryogenic Helium as high as $250~g/s$. Taking advantage of the very low kinematical viscosity of Helium at such a low temperature ($\nu \simeq 8 10^{-8}~m^2/s$ 
at $T=4.5~K$ and $P=1.2~bar$), very large Reynolds numbers can be achieved (up to $R_{\lambda} \simeq 6000$). Longitudinal velocity fluctuations 
have also been measured with a dedicated super-conducting `hot-wire' anemometer located at the center of the acoustic measurement volume, on the jet 
axis, 50 nozzle diameters downstream. For the highest flow rate the mean velocity in the measurement volume was $\simeq 4.4~m/s$ and the rms 
velocity $\simeq 1.25~m/s$.\\ 
The main flow characteristics of the different experiments are summarized in table \ref{FlowData}, as well as the range of investigated spatial wave-vectors.

\begin{table*}
\caption[]{Turbulent flows characteristics}
\label{FlowData}
\begin{tabular}{ccccccccc}

\hline
Flow &  \multicolumn{1}{c}{$R_{\lambda}$}
  & \multicolumn{1}{c}{$\lambda$ ($mm$)} & \multicolumn{1}{c}{$\eta$} 
($\mu m$) &
\multicolumn{1}{c}{{$\frac{\ell_o}{L}$}} &
\multicolumn{1}{c}{$2\pi/q_{min} $}  & \multicolumn{1}{c}{$ 2\pi/q_{max}$}  \\
\hline
LEGI Jet   & 530 & 8.4 & 180 & 0.75 & $ \simeq 3.5 \lambda$ & $\simeq 
10 \eta$  \\
& 740 & 6.7 & 120 & '' & $ \simeq 4.5 \lambda$ & $\simeq 15 \eta$  \\
& 785 & 6.5 & 107 & '' & $ \simeq 5 \lambda$ & $\simeq 17 \eta$  \\
\hline
CERN Jet
& 3450 & 0.69 & 5.45 & 1,5 & $ \simeq 10 \lambda$ & $\simeq 1.2 \lambda$  \\
& 4750 & 0.51 & 3.40 & " & $ \simeq 15 \lambda$ & $\simeq 1.8 \lambda$  \\
& 6090 & 0.42 & 2.55 & " & $ \simeq 17 \lambda$ & $\simeq 2 \lambda$  \\
\hline
\end{tabular}
\end{table*}

\subsection{Experimental Set-Up}

Acoustic waves are generated and detected using Sell-type transducers (see e.g. \cite{Anke74}). The Sell tranducers
are electro-acoustic reciprocal transducers, consisting in a circular plane piston, of diameter $L=15~cm$
and $L=5~cm$, respectively in the LEGI and the CERN jet flows, polarized with a high static voltage ($\approx 200~Volts~
DC$). The circular membrane made of a very thin (thickness $\simeq 15~\mu m$) mylar sheet achieves a large frequency 
band-width (between $1~kHz$ and $200~kHz$ in air flows). Indeed, when the working fluid is air, frequencies
higher than $200 kHz$ are strongly attenuated by dissipative process (viscosity and thermal conductivity). At small
acoustic frequencies (below $1~kHz$), the main frequency limitations stem from the turbulent noise generated by the
turbulent flow under investigation and the beams divergency due to diffraction effects. The transmitter is driven
by a MOSFET power amplifier (NF 4005, 100VA), while the acoustic pressure signal collected on the receiver is
converted into a voltage signal using a home made linear, low noise charge amplifier.\\ 
In all experiments, the scattering angle $\theta_{scat}$ is kept at a constant value (typ. $60 ^o$ in air flows) chosen so as to realize a trade-off between 
optimizing the sensitivity of the  measurements and limiting the spurious effects induced by diffraction side-lobes. 
Moreover, in any case, the size $L$ of the transducers is such that the linear size $({V_{scat}})^{1/3}$ is of the same order 
than the integral length scale of the turbulent flow (note that in the CERN experiment, the sound velocity is about $110~m/s$, 
nearly three times smaller than in the air at usual temperature). 
In a typical sound scattering experiment time series of the acoustic pressure signal are collected by the Sell
receiver and then sampled and recorded using Agilent E1430 VXIbus-based analog to digital converters with a high
precision (23 bits). Each E1430 module is provided with an analog anti-aliasing filter, digital filtering and decimation 
circuits and its own local oscillator allowing real-time heterodyne demodulation. In order to reduce phase noise and 
frequency drift, the sampling clock, the local oscillator clocks of the E1430 digitizer and the
clock of the frequency generator (Agilent E33120) which generates the continuous sinusoid signal driving the
transmitter are locked on the same 10 MHz master clock. The frequency of the local oscillator used for the numerical
demodulation by the E1430 digitizer is thus precisely tuned to the frequency $\nu_o$ of the incoming sound wave. 


\section{Spectral Analysis and Spatial Enstrophy Spectrum}

\subsection{Temporal and Spectral Caracteristics of the Scattered Pressure Signal}

We consider in this section, the averaged statistical spectral properties of the demodulated scattered pressure signal, recorded on
the detector for fixed incoming sound frequency $\nu_o$ and scattering angle $\theta_{scat}$, defining a single analysing wave-vector
${\bf q}_{scat}$ according to Eqns. (\ref{eqn_Diff}, \ref{q_scat}). Indeed, being a direct image of the time evolution of
$\Omega_{\perp}({\bf r},t)$ which is a stochastic signal, one expects the demodulated scattered pressure signal $z_{scat}(t)$ to be
also a complex random signal. We can express the complex signal $z_{scat}(t)$ as $\rho(t)e^{i\phi(t)}$, where
$\rho(t)$ (resp. $\phi(t)$) is the instantaneous amplitude modulation (resp. phase modulation). Typical evolutions of these two
quantities are represented on figure (4), obtained in the LEGI jet for a scattering wave-vector ($q_{scat}\lambda = 4.93 $).  
The upper plot shows the evolution, along time, of the instantaneous intensity $I(t) = |z_{scat}|^2 = \rho(t)^2$,
proportional to the instantaneous spatial enstrophy $|\Omega({\bf q}_{scat},t)|^2$ which clearly exhibits very large fluctuations. 
The lower part of figure (4) sketches the time evolution of the instantaneous phase $\phi(t)$. In this representation, the phase $\phi(t)$ has been
unwrapped, using an appropriate algorithm, to take into account phase jumps greater than $\pi$. The behaviour of the phase shift is roughly linear
with a mean slope $\frac{d\phi(t)}{dt} = 5800~rad/s$. 
Indeed, since in this experiment the direction of the scattering wave-vector ${\bf q}_{scat}$ has been chosen to be aligned with the mean flow 
velocity vector ${\bf V}_{avg}$, the linear phase shift is a mere consequence of a Doppler effect $\delta\nu(t) = \nu(t)-\nu_o = \frac{1}{2\pi}\frac{d\phi(t)}{dt}$ with :

\begin{equation}
\frac{d\phi(t)}{dt} = {\bf q}_{scat}{\bf V}_{avg}
\label{DopplerShift}
\end{equation}

This Doppler effect is related to the advection of the vorticity distribution by the large scale velocity field. Note however,
that local fluctuations of the instantaneous frequency shift are still visible, owing to turbulent fluctuations in time of the local velocity
itself. From the knowledge of ${\bf q}_{scat}$ ($739~m^{-1}$) and the estimated slope of  $\frac{d\phi(t)}{dt}$, we get an estimation of the 
local mean flow velocity : $7.79~m/s$ of the order of the mean flow velocity obtained with usual  hot-wire measurements, averaged on a much longer time ($7.39~m/s$). 

\begin{figure}[h] 
\vspace{6pc}
\centerline{
\includegraphics[width=12cm]{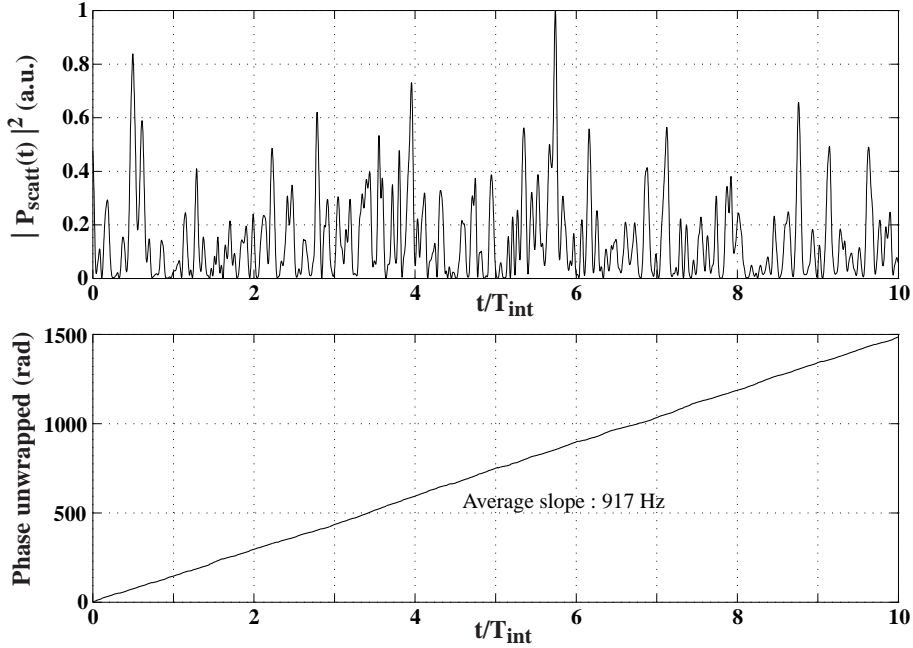}
\caption[]{Instantaneous scattered pressure intensity (upper) and phase modulation (lower) in the LEGI jet flow ($R_{\lambda} \simeq 785$)}
}
\label{TimeSignalsJetFlow}
\end{figure}

Let's turn now to the spectral properties of the demodulated scattered pressure signal. From a long time serie ($2^{20}$ points sampled at $F_s =
16384~Hz$, corresponding to $2500$ integral times $T_{int}$), we estimate the average Power Spectral Density (the spectrumhereafter will be referred 
as $PSD_{scat}(\delta\nu)$) using the usual Welch's averaged periodogram method \cite{Bendat_Piersol} (with a hanning window of length $2048$ samples 
and an overlap between consecutive segments of 1024 samples). The resulting PSD estimation, for the previous signal is plotted on figure (5). 
The spectrum $PSD_{scat}(\delta\nu)$, is asymmetric (with respect to the incident frequency $\nu_o$ (corresponding to a null Doppler frequency $\delta\nu=0$)
and has a Gaussian shape centered on a positive Doppler shift frequency as we choose ${\bf q}_{scat}{\bf V}_{avg} > 0 $. Note that, thanks to
complex nature of the demodulated signal $z_{scat}(t)$, we are sensitive to the direction of the flow: we are thus able to discriminate positive
(i.e. downstream) and negative (i.e. upstream) velocities. Indeed, the Gaussian shape of the scattered pressure $PSD_{scat}(\delta\nu)$ is a direct
consequence of the Gaussian statistics of the large scale velocity. Using a non linear Gaussian fit (the dashed line on figure (5)) : 

\begin{equation}
PSD_{scat}(\delta\nu) = \frac{A(\nu_{o})}{\sqrt{2\pi}\delta\nu_{rms}}exp^{-\frac{(\delta\nu - \delta\nu_{avg})^2}{2(\delta\nu_{rms})^2}}
\label{GaussianFit}
\end{equation}

we find a value $\delta\nu_{avg}~=~865.5~Hz$ and a value $\delta\nu_{rms}~=~224.5~Hz$ measuring  the width of the fluctuations of the advection velocity
around its mean value. The estimated values $\delta\nu_{avg}$ and $\delta\nu_{rms}$ can be converted into values of the advection velocity according to 
Eqn. (\ref {DopplerShift}) : $V_{avg} = 7.37~m/s$ and $V_{rms}~=~1.91~m/s$ in very good agreement with hot-wire anemometry measurements of the longitudinal 
velocity ($7.39~m/s$ and $1.89~m/s$).\\

\begin{figure}[h] 
\vspace{6pc}
\centerline{
\includegraphics[width=12cm]{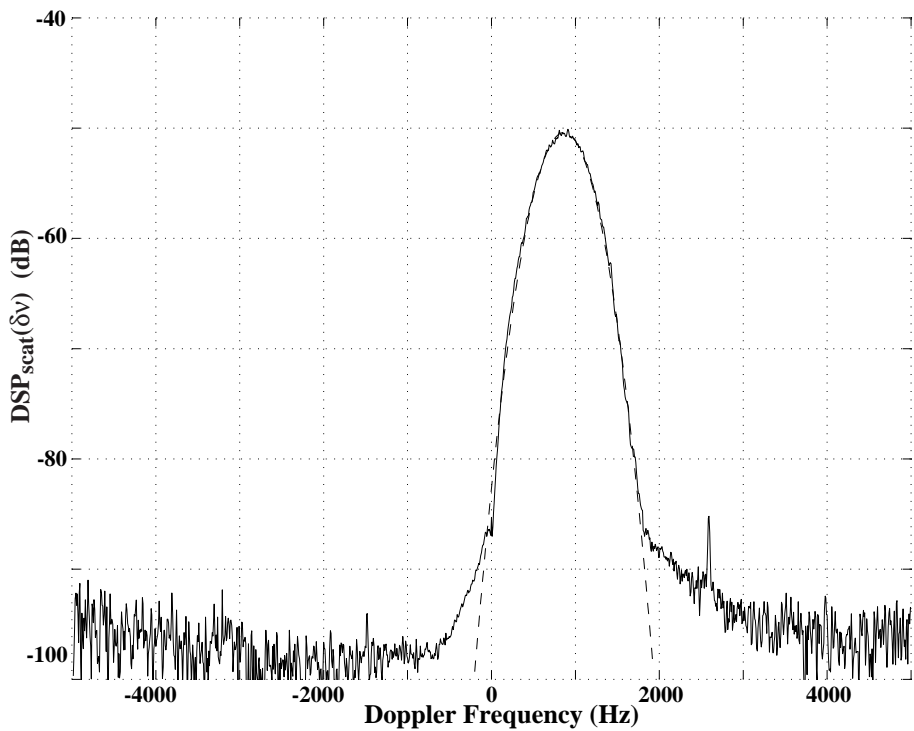}
\caption[]{Power Spectral Density of $\Omega_{\perp}({\bf q}_{scat},t)$ at a scale $\sim \lambda$ in the LEGI jet ($R_{\lambda} \simeq 785$)}
}
\label{JetFlowPSD}
\end{figure}

\subsection{Power spectral density of the vorticity Fourier modes in high Reynold numbers turbulence}

Acoustic scattering measurements have been also conducted in the CERN jet. Indeed, taking advantage 
of the very low kinematical viscosity of gazeous Helium at low temperature (around $4~^{\circ}K$), it is possible to achieve 
large Reynolds numbers, in a well controled flow (implying statistical stationnarity of the turbulent flows under investigation), 
while keeping moderate Mach number values in order for the Born approximation used in acoustic scattering analysis to apply. 
Figure (6) displays the PSD of the scattered acoustic pressure signal for $R_{\lambda}~=~3450)$. For this figure, the 
scattering angle is set to $\theta_{scat}~=~30^{\circ}$ and the incoming sound frequency to $\nu_{o}~=~110~kHz$ giving a scattering 
wave-number $q_{scat}~=~3250~m^{-1}$ equivalent to an anylsing scale $2.8 \lambda$. 

\begin{figure}[h] 
\vspace{6pc}
\centerline{
\includegraphics[width=12cm]{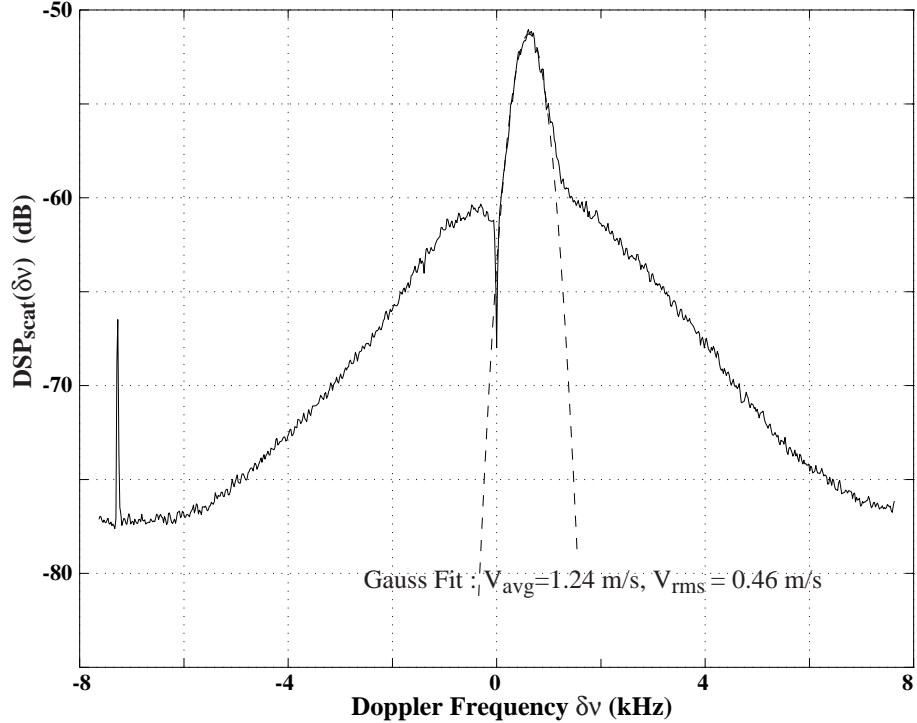}
\caption[]{Power Spectral Density of $\Omega_{\perp}({\bf q}_{scat},t)$, at a scale $\sim 3\lambda$ in the CERN helium jet flow ($R_{\lambda} \simeq 3450$).}
}
\label{CERNJetFlowPSD}
\end{figure}

As in the LEGI jet, a large amplitude scattered peak is still 
visible on the PSD; when fitted with a Gaussian, one gets an estimation of the mean advection velocity of $1.24~m/s$, in agreement 
with longitudinal velocity statistics obtained by hot-wire anemometry. Notice, however, that from the variance of the Doppler shift 
frequency $\delta\nu_{rms}$ we obtain an estimation of the variance of the velocity significantly larger than the one given by hot-wire 
anemometry.  The observed discrepancy, of order $20 \%$, could be ascribed to the finite size of the transducers. \\
A salient feature of the PSD at large Reynolds number, lies in the presence of significative scattered pressure amplitude at high Doppler frequency 
shifts, manifesting as exponential tails on both side of the principal Gaussian peak. Note that a close look at figure (5), 
also reveals a similar behavior, albeit with a much smaller amplitude (relative to the Gaussian peak). We have checked that the two 
exponential-like wings are approximately symetric with respect to the mean Doppler shift frequency, indicating that this high frequency part 
of the PSD is associated with vorticity events, advected by the large scale flow. As the typical frequencies involved in these exponential 
wings are an order of magnitude larger than the Doppler shifts, we are inclined to ascribe them to a much richer dynamic of the 
time evolution of the small scale vorticity events. Comparisons with PSD obtained at still  higher Reynolds numbers in the CERN experiment 
(up to $R_{\lambda}~=~6090$), indicate a clear enhancement of this high frequency behaviour (with  respect to the low frequency Gaussian part).

\subsection{Spatial Enstrophy Spectrum}

In the LEGI Jet experiment, the scattering measurements have been repeated for various incoming sound frequencies (the scattering angle being held to the constant value
$60~deg$) in order to scan a large range of spatial wave-vector numbers. 
For each incoming sound frequency $\nu_{o}$, the time spectrum $PSD_{scat}(\delta\nu)$ has been fitted with a Gaussian function (according to Eqn. (\ref {GaussianFit})) 
in order to estimate $\delta\nu_{avg}(q_{scat})$, $\delta\nu_{rms}(q_{scat})$ and $A(q_{scat})$. Figure (7) shows a plot of the evolution of
$\delta\nu_{avg}(q_{scat})$ and $\delta\nu_{rms}(q_{scat})$ with the wave-number $q_{scat}$. A linear evolution is roughly evidenced, according to Eqn. (\ref{DopplerShift}).

\begin{figure}[h] 
\vspace{6pc}
\centerline{
\includegraphics[width=12cm]{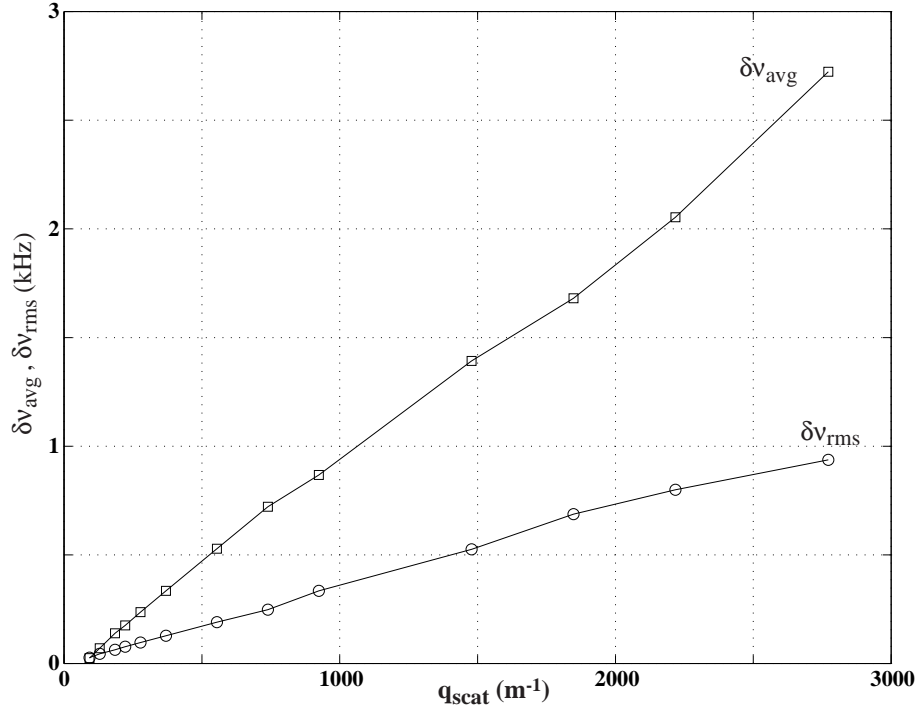}
\caption[]{Evolution of the mean $\delta\nu_{avg}$ and of the variance $\delta\nu_{rms}$ of the Doppler shifts, with the scattering
wave-number $q_{scat}$ in the LEGI jet ($R_{\lambda} \simeq 785$)}
}
\label{DopplerJetFlow}
\end{figure}

By integrating the scattered pressure PSD with respect to the Doppler shift frequency $\delta \nu$, one gets for each wave-number $q_{scat}$, a measure of the spatial enstrophy
spectrum according to the Wiener-Kintchine theorem \cite{Papoulis} : 

\begin{equation}
\mathop {\lim }\limits_{T\to \infty }{1 \over T}\int\limits_{-T/2}^{+T/2} {\left| {\Omega \left(
{q_{scat},t} \right)} \right|^2dt=\int\limits_{-\infty }^{+\infty } {\left| {\Omega \left( {q_{scat},\delta
\nu } \right)} \right|^2d(\delta \nu ) }}
\label{EnstrophyPSD}
\end{equation}

Using Eqn. (\ref{eqn_Diff}) and Eqn. (\ref{GaussianFit}): 

\begin{equation}
\int\limits_{-\infty }^{+\infty } {\left| {z_{scat}\left({\delta \nu } \right)} \right|^2d(\delta \nu )=A(\nu _o)}
\end{equation}

one sees that the quantity ${q_{scat}}^2 A(\nu _o)$, is a direct estimation of the spatial enstrophy spectrum $Enstro(q_{scat}) = {q_{scat}}^2
\left\langle {\left| {\Omega (q_{scat},t)} \right|^2} \right\rangle _t$ at wave-number $q_{scat} = \frac{4\pi\nu_o}{c}\sin(\frac{\theta_{scat}}{2})$ 
up to a multiplicative factor  $|H(\nu_{o})|^2$ which accounts for the frequency response of both transmitter and receiver around the working
frequency $\nu_{o}$. Actually, the transfer function $H(\nu_{o})$ is evaluated from measurements of the the electrical response delivered by the
receiver when the transmitter is driven by a wide-band electrical white noise. We have plot on figure (8) the variations of 
$Enstro(q_{scat})$ against the turbulent wave-numbers $q_{scat}$, for various length scales in the inertial range ($R_{\lambda}~=~785$). 

\begin{figure}[h] 
\vspace{6pc}
\centerline{
\includegraphics[width=12cm]{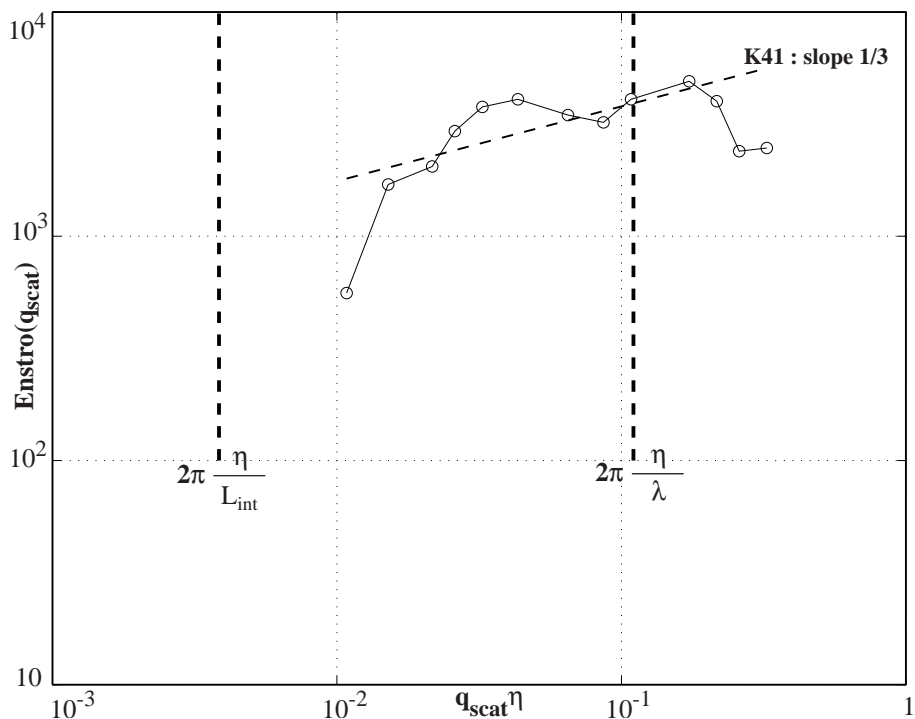}
\caption[]{Spatial enstrophy spectral density as a function of the  non dimensionnal wave-number in the LEGI jet ($R_{\lambda} \simeq 785$).}
}
\label{EnstroJetFlow}
\end{figure}

This log-log representation suggests a power-law scaling, with an exponent slightly positive and close to Kolmogorov 1941 prediction (dashed line with a slope $1/3$). 
Resorting to a scale invariance argument and assuming the statistical homogenity and isotropy of the flow, one expects (following Kolmogorov) the spatial enstrophy
spectrum to scale as : $Enstro(k) \propto {k}^2E(k)$ where $E(k) \propto k^{-5/3}$ is the turbulent energy spectral density. This latter result is the mere
consequence  of the definition of the vorticity field ${\bf \Omega}({\bf r}, t) = \nabla \wedge {\bf u}({\bf r}, t)$ which leads to 
${\bf \Omega}({\bf k},t) = -i{\bf k} \wedge {\bf u}({\bf k}, t)$ in the Fourier space. \\  
The spatial enstrophy spectrum differs significantly from time-spectra measured with multi-wires velocity probes \cite{Wallace86, Tsin92, Anton97}. In its principle, 
the acoustic scattering measurement gives a good resolution in the Fourier space, thus it gives the true 1D spectrum, as the spatial Fourier transform of the vorticity 
field is evaluated at a well defined spatial {\em wave-vector}. In the case of multi-wires anemometry measurements of the Eulerian vorticity, the vorticity spectrum 
evaluated at {\em  wave-number} $q_x = 2\pi \frac{\nu}{V_{avg}}$ involves the contributions  of all the spatial wave-vectors with the 1D projection $q_x$ \cite {Tennekes, Batch53}. 
Taken as a whole, such a `1D spectrum' is mainly dominated by the small scales (large $q_x$) viscous cut-off and hence does not exhibit any inertial scaling \cite{Pumir2001}.

\subsection{Time-Frequency Distributions}

A detailed examination of the time evolution of the scattered pressure signal (phase and amplitude as depicted on figure (4)), 
reveals strong fluctuations which can be ascribed to the passage in the measurement volume of finite duration vorticity events. In turn the scattered acoustic
signal exhibits a time-varying spectrum, as the frequency of the measured pressure signal drops from $\nu_0$ to $\nu_0 + \delta\nu$ when a vorticity event is
present, with significative energy at the wave-vector ${\bf q}_{scat}$ in the spatial Fourier domain. Due to the randomness of the advection velocity of the
vorticity field, the true vorticity spectrum is blurred and wiped out by the time averaging process. 
Thus we seek after a representation of the signal which preserves simultaneously the time and frequency information.\\  
Time-Frequency distributions \cite{Flandrin99,Will92,Baudet99}, among which the Wigner-Ville distribution (WVD) is the most famous, provide a simultaneous
description (in time and frequency) of the energy distributions of non-stationary signals in the time-frequency plane. 
The Wigner-Ville distribution $W_z(t,\nu)$ of a complex analytic signal $z(t)$ is defined as the time Fourier transform of the local auto-correlation function of $z(t)$:
$R_z(t,\tau)~=~z(t+\frac{\tau}{2})z^{*}(t-\frac{\tau}{2})$ with respect to the lag variable $\tau$:

\begin{equation}
W_z(t,\nu)=F_{\tau}[z(t+\frac{\tau}{2})z^{*}(t-\frac{\tau}{2})]=F_{\tau}[R_z(t,\tau)]
\label{WVD_Def}
\end{equation}

where $z(t)$ is the time signal and $z^{*}(t)$ its complex conjugate. The operator $F_{(.)}$ denotes the Fourier transform operator, with respect to the variable $(.)$. 
This definition generalizes, in some way, the Wiener-Kinchine theorem to non-stationary signals, for which the auto-correlation function depends on two time variables: 
$R_z(t_1,t_2)=\left\langle {z(t_1)z*(t_2)} \right\rangle \ne R_z(t_1-t_2)$. 
From an experimental and heuristic  point of view, the effect of the operator $R_z(t,\tau)$ is equivalent  to a local phase conjugation of  the signal which enhances 
the phase derivative related to the Doppler shift \cite{Flandrin99}.  Similarly, one usually defines the ambiguity function $A_z(\theta,\tau)$ as the inverse Fourier 
transform ($F_{t}^{-1}$) of $R_z(t,\tau)$ with respect to the first variable t.

\begin{equation}
A_z(\theta,\tau)=F^{-1}_t[z(t+\frac{\tau}{2})z^{*}(t-\frac{\tau}{2})]=F_{t}^{-1}[R_z(t,\tau)]
\end{equation}
Thus, $W_z(t,\nu)$ and $A_z(\theta,\tau)$ are related by the two-dimensional Fourier transform
\begin{equation}
W_z(t,\nu)=\int \int A_z(\theta,\tau) e^{-j(t\theta + \nu \tau)}d\theta d\tau
\label{AF_Def}
\end{equation}
The ambiguity function reduces to a deterministic correlation function in time if $\theta$ is set to zero. Similarly, as can be seen in a dual form which 
starts with $Z(\nu)~=~F_{t}[z(t)]$, it  can be seen as a deterministic correlation of spectra if $\tau$ is set to zero. Starting from the Wigner-Ville distribution
$W_z(t,\nu)$,  a wide class of Time-Frequency Distributions (TFD) can be developped by means of bi-dimensional smoothing operations with appropriate kernels
$\phi(\theta,\tau)$:

\begin{equation}
C_z(t,\nu, \phi)=
\int \int \int e^{j((\xi-t)\theta- \nu \tau)}\phi(\theta,\tau)z(\xi+\frac{\tau}{2})z^{*}(\xi-\frac{\tau}{2})d\xi d\tau d\theta
\label{CohenClass_DefF}
\end{equation}

The purpose of such a generalisation of the WVD is to provide the experimentalist with an operationnal definition of the Time-Frequency Distribution $TFD_{z}(t,\nu)$ of any
signal $z(t)$, which can be interpreted as a time-frequency power density: $TFD_{z}(t,\nu)\Delta \nu \Delta t$ is the energy of the signal $z(t)$ between the instants $t$ and
$t+\Delta t$, in the frequency band $[\nu~\nu+\Delta \nu]$.
Indeed, one can show that with such a definition (smoothing kernels), the so-called covariance property (invariance with respect to time and frequency translations or
modulations) and marginal preservations are respected : 

\begin{eqnarray}
PSD_{z}(\nu) & = & \mathop {\lim }\limits_{T\to \infty }{1 \over T}\int\limits_{-T/2}^{+T/2}  TFD_{z}(t, \nu)~dt\\
|z(t)z^{*}(t)| & = & \int\limits_{-\infty}^{+\infty}  TFD_{z}(t, \nu)~d\nu
\label{TFT_Margin}
\end{eqnarray}

\begin{figure}[h] 
\vspace{6pc}
\centerline{
\includegraphics[width=12cm]{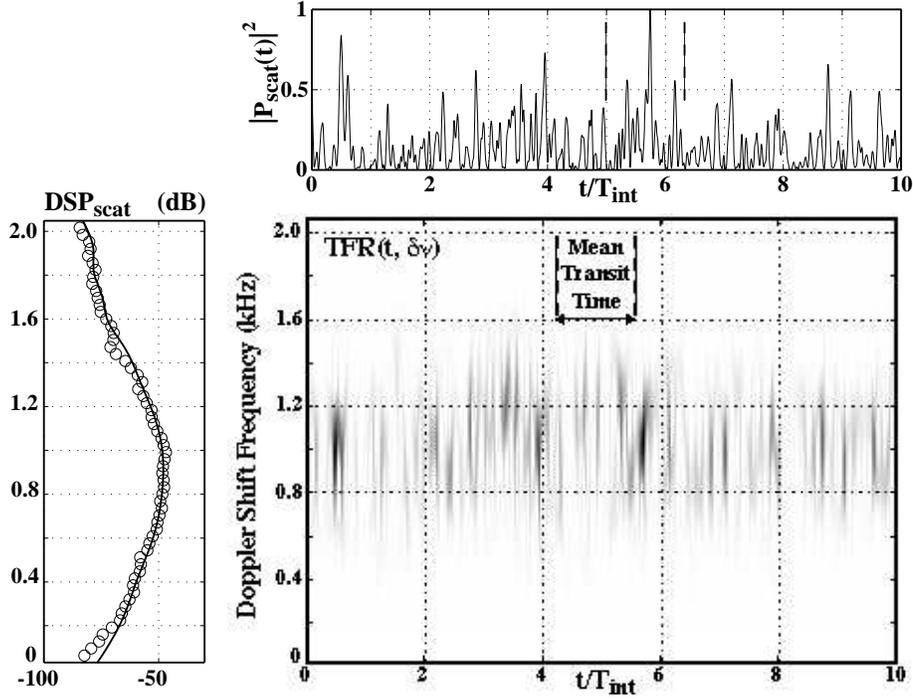}
\caption[]{Time-Frequency Analysis of $z_{scat}(t)$ in the LEGI air jet flow ($R_{\lambda} \simeq 785$). Center : $TFD_{z}(t, \nu)$, Left : PSD marginal (o) and averaged PSD (-), 
Top : instantaneous power marginal}
}
\label{TFRSig_SpectJetFlow}
\end{figure}

An example of such a Time-Frequency analysis is displayed on figure (9) corresponding to the time-serie presented on figure
(4). The center part of figure (9) is a pseudo-color image (with a linear gray-scale) of the
level $TFD_{scat}(t,\nu)$. This representation evidences the existence of finite duration events, with local random Doppler shifts corresponding to their own advection
velocity. On the left side of figure (9), the time averaged spectrum has been represented, together with the frequency marginal $PSD_{z}(\nu) = 
\mathop {\lim }\limits_{T\to \infty }{1 \over T}\int\limits_{-T/2}^{+T/2}  TFD_{scat}(t, \nu)~dt$ of $TFD_{scat}(t,\nu)$. The latter representation demonstrates the blurring
effect of the random advection by the large scale velocity field. A way to overcome this random sweeping effect is to account for the instantaneous advection velocity 
(at each time where a vorticity event is present, detected by a large amplitude of $\rho (t)$) . Noting that, as the major effect of the advection
velocity is to induce a phase modulation according to a Doppler frequency shift, we have observed that random advection velocity can be safely compensated and fully eliminated
by considering only the modulus $\rho (t)$ of the scattering complex pressure signal $z_{scat}(t)$ thus ignoring the phase information $\phi(t)$.


\section{Time-Space Correlations}

\subsection {Characteristic time of a vorticity mode}

Following the later remark, the characteristic time of any spatial vorticity Fourier mode $\Omega_{\perp}({\bf q}_{scat},t)$ can be studied by considering only 
its modulus $\propto \rho(t)$. Figure (10) shows the normalized auto-covariance $Cov_{\rho\rho}(\tau)$ of the vorticity modulus for 
different wave-vectors ${\bf q}_{scatt}$ aligned with the jet axis in the case of the LEGI experiment. The typical shape of these correlation curves (right figure)
brings out two times: a short one (hereafter refered as $\tau_s$) characterized by a fast decrease of the correlation, and a longer one (noted $\tau_\ell$) associated 
with low, but significant correlation values.  Qualitatively, the short time part can be  rather well fitted by a Gaussian curve (as shown in the left figure), while 
the long time correlation part has rather an exponential-like shape.

\begin{figure}[h] 
\vspace{6pc}
\centerline{
\includegraphics[width=12cm]{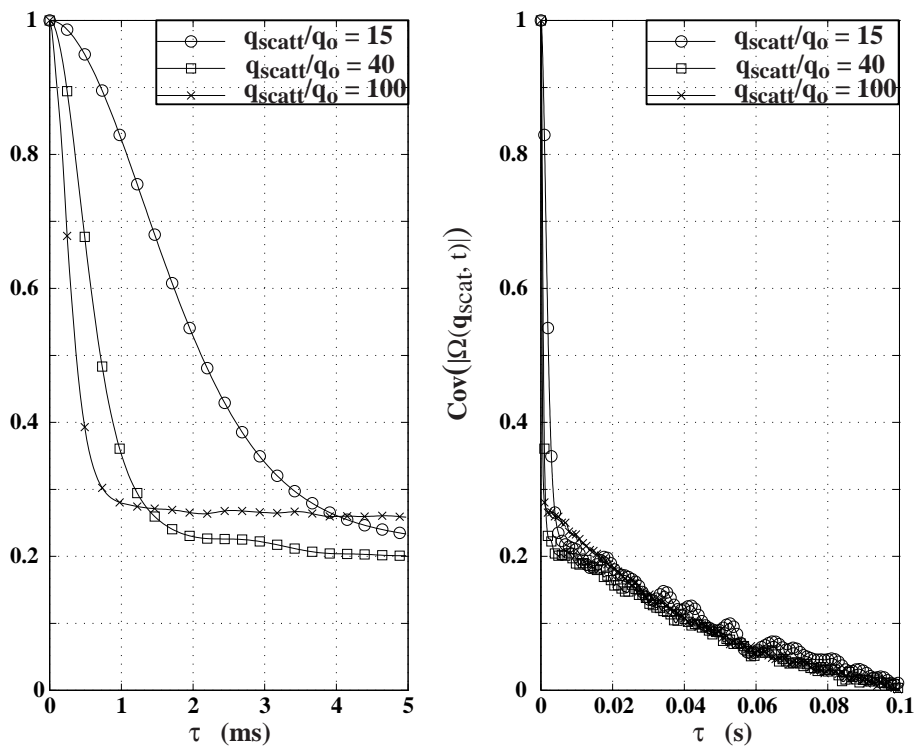}
\caption[]{Time correlation of vorticity modulus for different scales 
$1 / q_{scat}$ in the LEGI jet ($R_{\lambda} \simeq 740$) (log-lin coordinates in the insert)}
}
\label{Timecorrelation}
\end{figure}

The short time corresponds to the average time during which the component of a vorticity event $\Omega_\perp$ is tuned with
the analyzing wave-number $q_{scatt}$. This short time $\tau_s$ clearly varies with the analyzing scale $q_{scatt}^{-1}$ (see below).\\
In our experiments, the duration $\tau_s$ was systematically much shorter than the mean transit time in the measurement volume  evaluated from the size $L$ of 
the scattering volume and the mean advection velocity: $\tau_s \ll \frac{L}{V_{avg}}$ (the mean transit time in this experiment is $\simeq 25~ms$). 
Accordingly, this latter observation reveals the Lagrangian feature of  the time analysis performed on the spatial Fourier modes of vorticity by means of the covariance estimator
$Cov_{\rho\rho}(\tau)$. The short time corresponds to the average duration during which the component $\Omega_\perp$ of a vorticity event is observed at the
the analyzing wave-number $q_{scat}$. In our experiments, this duration $\tau_s$ is systematically shorter than the corresponding
crossing time in the measurement volume. This short time $\tau_s$  clearly varies with the analyzing scale $q_{scat}^{-1}$ .
On the contrary, the long time part of the correlation does not seem to depend on the scale $q_{scatt}^{-1}$. In particular, we observe that the correlation
curves cross the null value axis at a typical time of the order of $\tau_0 = 2 \pi (U_{rms} q_0)^{-1}$ where $q_0$ is the wave-number characteristic of the 
integral length scale of the flow. In our experimental conditions, this large scale time is longer than the crossing time of the measurement volume.
This weak auto-correlation of the modulus $\rho(t)$ at long time delays $\tau$ (larger than the duration time $\tau_s$) suggests that it should reflect the time behaviour 
of the inter-event dynamic (involving several isolated events). The significantly non zero values, is an indication of the organization of several events belonging to 
the same large structure of the turbulent flow. We will not investigate further these long time correlation property as they clearly do not correspond to a Lagrangian 
description of the flow.\\
Focusing on the short time $\tau_s$, the main question is to know how it depends on both the length scale and the Reynolds number.

\subsection {Scale dependence}

The figure (11) shows the behaviour of the short time $\tau_s$  with respect to the length scale $q_{scat}$. 
Here $\tau_s$ has been arbitrarily defined as the width at the half correlation value. We checked that another definition of $\tau_s$ (for instance, the standard 
deviation of the Gaussian fit) does not change the results. 

\begin{figure}[h] 
\vspace{6pc}
\centerline{
\includegraphics[width=12cm]{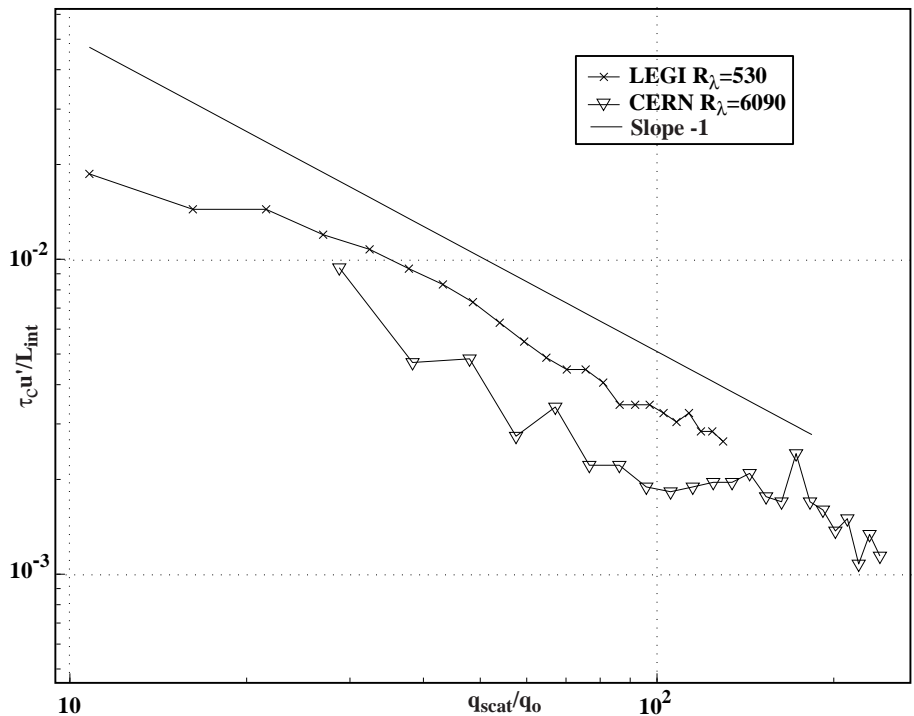}
\caption[]{Scaling of $\tau_s$ (normalized by the large scale time $\tau_0$)
with the wave number $ q_{scat}$: ($\times$) $R_{\lambda} \simeq 530$), ($\nabla$) $R_{\lambda} \simeq
6090$. }
}
\label{Sweeping_scaling}
\end{figure}

Despite the low extent of the analyzing wave-number range ($\sim$ one decade), data
are in better agreement with a slope $-1$ than a slope $-2/3$, meaning that the charateristic time of $ \Omega_{\perp} (\vec {q_{scatt}}, t)$, depends linearly on the 
scale, whatever the Reynolds number. Thus, $\tau_s$ does not scale as the eddy turn over time ($\sim q^{-2/3}$). For the smallest inertial scale range (the only one 
attainable with our acoustic scattering device), experiments show that $\tau_s$ is actually proportional to the local sweeping time, defined as 
$\tau_{sweep} = 2 \pi (U_{rms} q_{scat})^{-1}$.
Putting forward an analogy between a given Fourier mode $\Omega_{\perp}(q_{scatt)}$ (in the Fourier space) and a velocity increment $\delta v(\delta r)$ over a given separation
$|delta r \sim {q_{scatt}}^{-1}$ (in the real space) then, the experimental data of the figure (11) are in an agreement with those obtained, numerically by Sanada 
et al. \cite{Sanada92}, and experimentally by Xu et al. \cite {Xu01}. Such a result would suggest that $\tau_{sweep}$ is the pertinent time of the vortex stretchings and/or tippings 
due to the local velocity gradients. In both papers \cite{Sanada92,Xu01}, it appears that for the largest scales of the inertial range, the eddy turn-over time possibly becomes 
again shorter than the sweeping one. Unfortunately our acoustic device can not reach such inertial scales.

\subsection {Reynolds dependence}

The figure (11) also indicates that the ratio of the vorticity mode life-time $\tau_s$ on the local sweeping time $\tau_{sweep}$ weakly depends on 
the Reynolds number. 

\begin{figure}[h] 
\vspace{6pc}
\centerline{
\includegraphics[width=12cm]{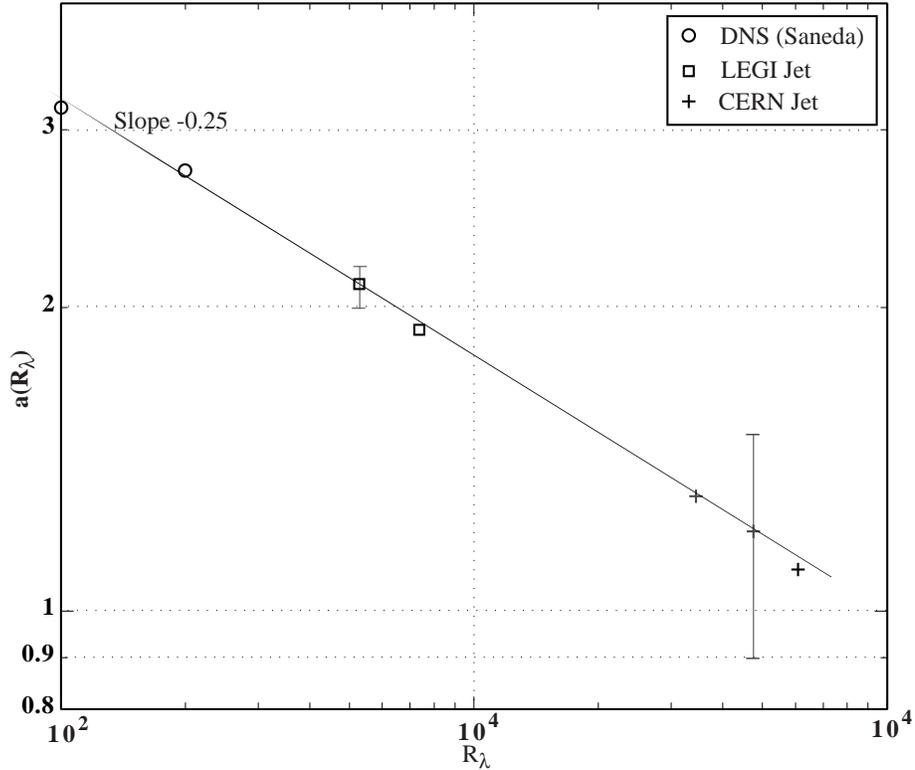}
\caption[]{Power law behaviour of $a(R_\lambda)$ in jets for $ 530 
\leq R_\lambda \leq 6090$}.
}
\label{Powerlaw}
\end{figure}

The figure (12) shows how this ratio $a(R_\lambda) = \frac {\tau_s}{\tau_{sweep}}$ behaves with the Reynolds number.
Thanks to the large range of investigated Reynolds numbers, ($100 \le R_\lambda \le 6000$), it is reasonable to infer a  power law scaling with an exponent
close to $-1/4$. Note that the numerical data of Sanada et al. \cite{Sanada92} are in very good agreement with our experiments (the two points of figure 12
have been taken from the figure 8 of \cite{Sanada92}).\\
In others words, for a fixed scale $2 \pi (q_{scat})^{-1}$, the vorticity correlation time decreases with the Reynolds number.
In connection with the vorticity time-correlation, Novikov \cite{Novikov93,Novikov94} has introduced a typical characteristic length $\ell_c$ which is the inertial scale 
where the  viscous dissipation of the enstrophy is balanced by the large scale effects. In terms of time scales, the Novikov prediction is such as $\tau_c \simeq  \tau_0
1.40 R^{-2/5}_\lambda$ (where $\tau_c$ is the time scale of the vorticity correlation).\\
This prediction can not be directly compared to the experimental scaling of the figure (12) because the latter relies on the analysis of a single length scale whereas the
theoretical result is based on the global budget of the vorticity correlations involving all the scales. However, one expects that experimental data are not 
in agreement with the previous prediction, because it is mainly derived from Kolmogorov arguments and it does not account for sweeping effects.


\section{Lagrangian Velocity Measurements}

The interest in the statistics of the Lagrangian velocity (following the fluid particules) in turbulent flows has been recently renewed by the emergence of new experimental 
techniques for the continuous  tracking along time of isolated particules \cite{Pinton01,Pinton02} (using acoustical devices) or the measurement of a large number of individual 
particule acceleration events \cite{Voth98} (using fast optical techniques). In the first  type of experiment \cite{Pinton01}, a very few number (ideally one) of solid beads 
are injected in a closed  turbulent water flow. 
The trajectory of a single particule is  then followed along time intervals, as large as the integral length scale of the turbulent flow, using an acoustic scattering device 
consisting  in an array of small transducers  (in a backscattering configuration).\\  
The purpose of the present experiment is to extend the latter acoustic tracking technique to the study of the Lagrangian turbulent velocity  in open turbulent air flows. 
Material particules, with an acoustic index different from the  acoustic index of the propagating medium, can efficiently scatter acoustic waves \cite{Pierce}.  
Provided the acoustic contrast (the relative difference between the acoustic  index of the particule and that of the medium) is sufficient, acoustic scattering  by particules 
injected in a turbulent flow can largely overcome the acoustic scattering by  the turbulent vorticity fluctuations. When a particule is advected by the local flow  velocity, 
the scattered pressure amplitude $p_{scat}(t)$, at a large enough scattering  wave-number ($q_{scat} r \geq 2\pi$, where $r$ is the typical  size of the particule), 
will exhibit Doppler shift phase modulations. 
We have seeded the LEGI jet, with very small soap bubbles  (of diameter $\leq 3~mm$), at a small rate  of injection (typ. $\leq 100~bubbles/s$) to have at most one particule at 
the same time in the scaterring volume $V_{scat}$.\\
The soap bubbles are inflated with gazeous helium,  with a home-made apparatus, so as to match pecisely the average density of the bubble with
the density of the surrounding fluid  (the low density helium compensates properly  the mass of the soap film). Thanks to this density matching, and  to the small size of the bubbles 
(limiting wake effects  behind the bubble \cite{Pinton01}), inertial effects are small and  each isolated particule is expected to follow, nearly  instantaneously, the fluid 
particule velocity.  In this experiment, the scattering angle is set to $120^{\circ}$ in order  to increase, both the wave-number $q_{scat}$  and the size of the scatterring 
volume $V_{scat}$ in the direction of ${\bf q} _{scat}$ (see figure (3)) up  to a length larger than the integral length scale $L_{int}$ of the turbulent 
flow. Moreover, at such a large scattering angle, the contribution of the vorticity fluctuations to  the scattering amplitude is  significantly reduced (see figure (2)).\\
In this preliminary experiment, the mean velocity was about $4~m/s$ ($R_{\lambda} \sim 530$).

\subsection {Scattered pressure signal PSD and covariance}

\begin{figure}[h] 
\vspace{6pc}
\centerline{
\includegraphics[width=12cm]{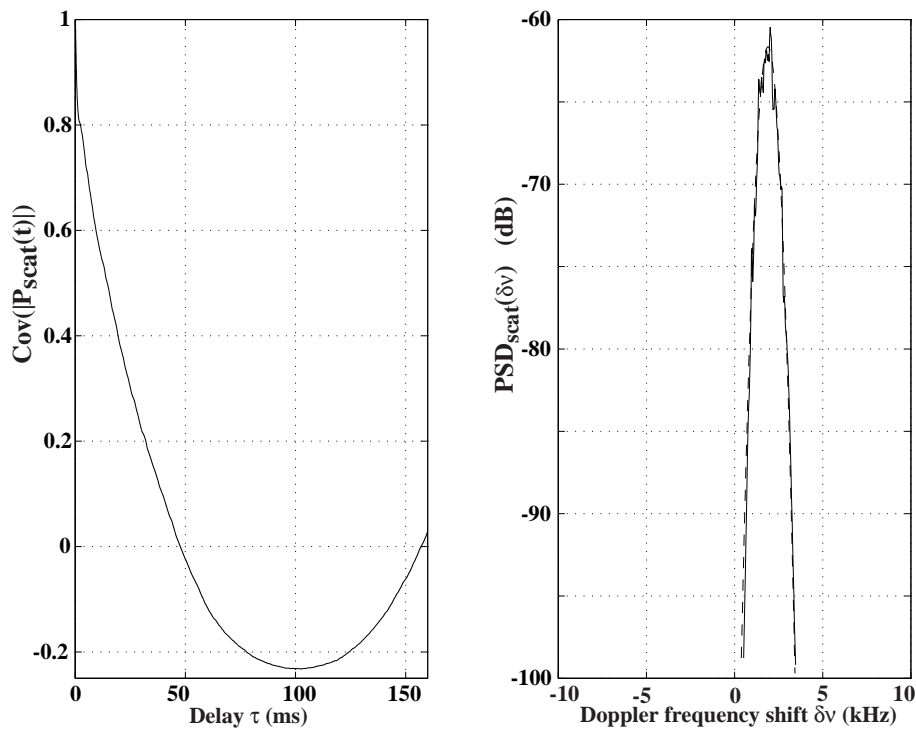}
\caption[]{Normalized Covariance $C_{\rho\rho}(\tau)$ of $\rho_{scat}(t)$ (left) and $PSD_{scat}(\delta \nu)$ (right) of the acoustic pressure scattered by soap bubbles in 
the LEGI turbulent jet flow at $R_{\lambda} \simeq 500$.}
}
\label{CovPsd_BubbleScatt}
\end{figure}

A first experiment has been performed to study the correlation and the spectrum of the acoustic pressure scattered by the soap bubbles. Simultaneous acquisitions have been realized with 
the superposition of two incident frequencies ($80~kHz-100~kHz$ and $100~kHz-120~kHz$). The two demodulated scattered signals around each incident frequency is then extracted by means of 
a band pass filtering and a demodulation. Such a double channel experiment allows a direct comparison of the detection at different incoming sound frequencies. In all experiments, the passage 
of a bubble in the scattering volume is simultaneously detected (around each $\nu_o$) as a strong increase of the scattered pressure intensity $I_{scat}(t) = \rho(t)^2$, indicating a large 
signal to noise ratio  of the acoustic scattering set-up. Also, the Doppler shift frequency, around each incoming frequency $\nu_o$ scales continuously in time with $\nu_o$ according to Eqn.
\ref{DopplerShift}. The high level of similarity at each time $t$, of both $\rho(t)$ and $\frac{d\phi(t)}{dt}$ is a good indication of the efficiency of the acoustic scattering
technique. Moreover, thanks to the small bubble flow rate, we effectively observe isolated bubbles passing through the scattering volume.\\ 
Indeed, the time interval during which the scattered pressure is significant is of order $\Delta T =\frac{\Delta L}{V_{avg}}$, where ${\Delta L}$ is the size of the scattering volume in the
direction of ${\bf q}_{scat}$ (aligned with the mean flow). This is illustrated by the plot of the normalized auto-covariance of $\rho(t)$ on figure (13-left). 
The covariance, decreases slowly from its maximum value (at the delay $\tau =0$), to a null value at the delay $\tau = \frac{\Delta L}{V_{avg}} \simeq 50~ms$. 
Remind that in the case of acoustic scattering by vorticity fluctuations, a much more rapid decrease of the auto-covariance is observed leading to an estimation of the duration of vorticity events
of a few $ms$ (cf figure (10)). The length of the negative part of the auto-covariance can be ascribed to the mean  time between the passage of two consecutive bubbles in the scattering volume. The
demodulated scattered pressure signals have been largely over-sampled at $F_s~=~32768~Hz$ and further  band-pass filtered with a Butterworth digital filter of order 8. The right part of figure (13)
displays the power spectral density for the scattered pressure signal  at an incoming frequency $\nu_o = 80~kHz$. 
As expected, the shape of $PSD_{scat}(\delta \nu)$ is Gaussian ($\delta \nu _{avg}$ and $\delta \nu _{std}$, in excellent agreement with the statistics of the longitudinal flow velocity (given by
hot-wire anemometry).

\subsection {Lagrangian velocity PDF and spectrum}

To investigate the statistical properties of the Lagrangian velocity, a very large number of bubble trajectories is required for purpose of statistical convergency. 

\begin{figure}[h] 
\vspace{6pc}
\centerline{
\includegraphics[width=12cm]{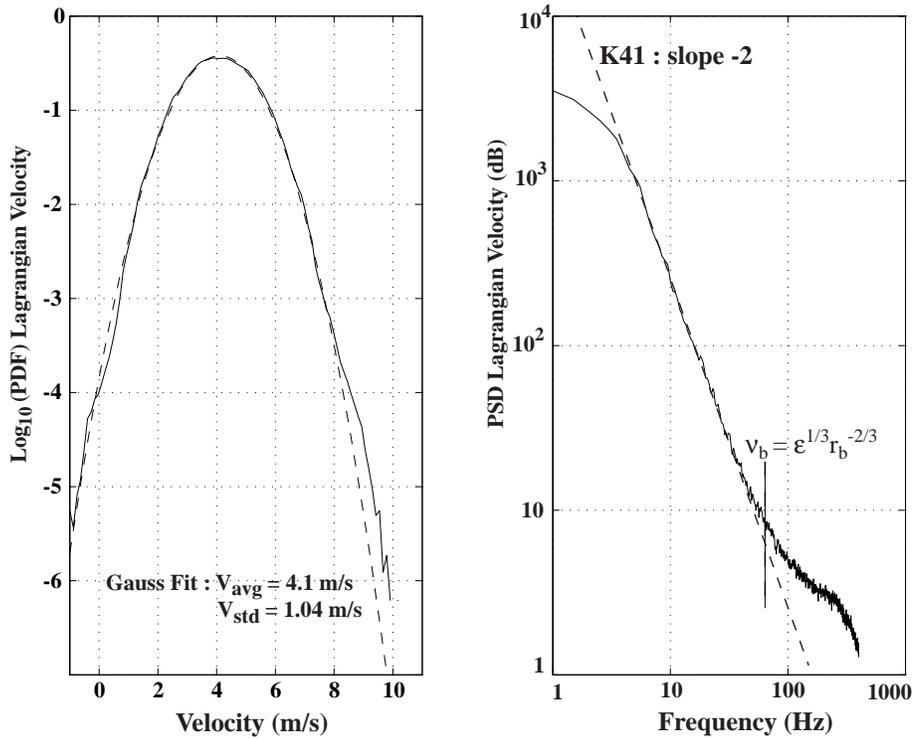}
\caption[]{PDF (left) and PSD (right) of the Lagrangian Velocity in the LEGI turbulent jet flow at $R_{\lambda} \simeq 500$}.
}
\label{PdfPsd_LagrangianVelocity}
\end{figure}

Thus we have performed a long time experiment, collecting $2.7~10^8$ complex data points sampled at $F_s~=~32768~Hz$. Using a post-processing algorithm, 
relying on an appropriate thresholding of the scattered acoustic intensity $I_{scat}(t)$, the passage of isolated bubbles in the scattering volume are 
identified and localized.\\ 
About $13~10^6$ passing bubbles have been identified and localized. After a band-pass filtering, the projection of the instantaneous velocity (onto the direction of the jet axis) 
of each isolated bubble is estimated  from the unwrapped phase signal $\phi(t)$ (according to $v_{bubble}(t)=\frac{1}{q_{scat}}\frac{d\phi(t)}{dt}$) with a finite difference schema.\\
From all these Lagrangian velocity signal we have computed firstly the probability density function $PDF(v_{bubble})$ of the Lagrangian velocity depicted on figure 
(14 left). The Lagrangian PDF has a nice gaussian shape, indicating a mean value and rms values of the Lagrangian velocity (note that they are ensemble 
averages) close to the corresponding Eulerian values (determined with hot-wire anemometry). The equality of the Lagrangian and Eulerian probability densities is a well 
known property of homogeneous turbulent flows \cite{Monin&Yaglom,Pope2000}.\\
The time power density spectrum $PSD_{Lag}(\delta \nu)$ have been computed with the Welch's averaged periodogram method. 
The resulting PSD is displayed on figure (14 right). In this log-log representation, a clear power-law scaling over roughly one  decade is visible. 
The exponent of the observed power law scaling is closed to the value $-2$ in agreement with the theoretical prediction of Kolmogorov \cite{Monin&Yaglom}. 
At higher  frequencies, the DSP display a significant cut-off, above the frequency $\nu_c \simeq 70~Hz$ very close to the low-pass filtering frequency due to 
the radius  $r_{b} = 3~mm$ of the bubbles ($\nu_{b} \simeq \epsilon^{1/3}r_{b}^{-2/3} \simeq 70~Hz$ where $\epsilon$ is the mean energy dissipation rate).


\section{Concluding remarks}

Wave scattering (e.g. light and neutron scattering in condensed matter and X-ray scattering in solid state physics) has proved to be a powerful tool in the study and of complex systems (e.g.
phase transitions, DNA structure and denaturation). Turbulent flows are also known to be very complex and disordered.  The scattering of coherent acoustic waves by turbulent vorticity fluctuations
allows the direct and continuous (in time) probing of a spatial Fourier mode $\Omega_{\perp}({\bf q}_{scat},t)$, at a well defined  spatial wave-vector ${\bf q}_{scat}$. In addition, the  acoustic
scattering technique is non intrusive and sensitive to the local orientation of the vorticity field \cite{Baudet91}.  Thus it appears as a relevant means of investigation of turbulence, where
the vorticity dynamic is known to play a crucial role (e.g. vortex stretching and bending), particularly  in the energy transfers across the length scales \cite{Dubrulle03, Nazarenko03}. By
selected different spatial wave-vectors, one is able to analyse scale by scale the statistical properties of the turbulent vorticity distribution.  For example, we have evidenced a scaling law with
exponent $1/3$ for enstrophy spectral density, in agreement with the Kolmogorov prediction. Time correlation of the  amplitude (modulus) vorticity modes reveals a short life-time of the vorticity
events at a fixed scale (possibly coherent structures \cite{Baudet99}). As the average duration of the detected vorticity events is much smaller than their mean transit time across the scattering
volume we argue that our analysis is of a  Lagrangian type. Still, the vorticity correlations also exhibit a long time behaviour, with a lower level of correlation, indicating probably a large scale
structure of the vorticity distribution which could be interpreted as the consequence of the turbulent cascade process.\\
Acoustic scattering can also be used to access Lagrangian statistics of the turbulent velocity thanks to the possibility to detect very small particules (using high enough ultrasound frequencies),
on very large volumes (of order the integral length scale of the flow). With this preliminary experiment, we have found again, in an open flow, Lagrangian probability density function of the
turbulent velocity  field equal to the Eulerian PDF as expected. Beside, we have also evidenced a Lagrangian power law scaling ($\propto \omega^{-2}$) for spectral power
density, according to the Kolmogorov 41 theory. The power law scaling extends over roughly one decade. Further investigation are in progress. Inparticular, we have already generated and successfully
detected bubbles with a smaller radius ($\leq 1~mm$) and thus we expect to increase the latter power law scaling  range towards the dissipative time-scales.

\acknowledgements
We acknowledge financial support from R\'egion Rh\^one-Alpes (Programme Th\'e\-ma\-ti\-ques Prioritaires `Acoustique', Contract Number 01-867-801 and -802),
the French Minist\`ere de l'\'Education Nationale et de la Recherche (DSU~2) and the Universit\'e Joseph Fourier-Grenoble I. We are also indebted to P. Lebrun 
and O. Pirotte for their valuable help in running the CERN experiment.\\
We are thankful to Bob Antonia for his contribution to this work. In the nineties, he warmly motivated us to seek after an alternative tool for the study of vorticity in turbulent flows.

\end{article}

\begin{thebibliography}{99}

\bibitem{Frisch95} Frisch, U.: 
\newblock {{\em Turbulence}},
\newblock {Cambridge University Press, 1995.}

\bibitem{Monin&Yaglom} Monin, A.~S., and A.~M. Yaglom: 
\newblock {{\em Statistical Fluid Mechanics}},
\newblock {The MIT Press, Third edition, 1985.}

\bibitem{Goodman} Goodman, J.~W.:
\newblock {{\em Statistical Optics}}, 
\newblock {Wiley-Interscience, 1985.}

\bibitem{Pierce} Pierce, A.~D.:
\newblock {{\em Acoustics}}, 
\newblock {Acoustical Society of America, 1989.}

\bibitem{Morse} Morse, P.~M., and K.~U. Ingard:
\newblock {{\em Theoretical Acoustics}}, 
\newblock {Princeton University Press, 1986.}

\bibitem{Engler89} Engler R. \& al.: 1982 
\newblock {{\it J.Acoust. Soc. Am.} {\bf71} (1), pp. 42--50.}

\bibitem{Kov76} Ho, C.~M., and \& L.S.G. Kov\`asznay L.~S.~~G.: 1976,
\newblock {`Propagation of a coherent acoustic wave through a turbulent shear flow',}
\newblock {{\it J.Acoust. Soc. Am.} {\bf 60} p.~40--45.}

\bibitem{Korm80} Korman M.~S., and Beyer R.~T.: 1976, 
\newblock {`The scattering of sound by turbulence in water'}
\newblock {{\it J.Acoust. Soc. Am. } {\bf 67} (6) pp.~1980--1987.}

\bibitem{Obuk53} Obukhov, A.~M.: 1953,
\newblock {`Effect of weak inhomogeneities in the atmosphere on sound and light propagation',}
\newblock{{\it Izv. Akad.Nauk. Seriya Geofiz. } {\bf 2} pp.~155--165.}

\bibitem{Kraic53} Kraichnan, R.~H.: 1953,
\newblock {`The scattering of Sound in a Turbulent Medium',}
\newblock{{\it J.Acoust. Soc. Am. } {\bf 25} pp.~1096--1104.}
 
\bibitem{Chu58} Chu B.~T., and L.~S.~G.~Kov\`asznay: 1958,
\newblock {`Non-linear interactions in a viscous heat-conducting compressible gas',}
\newblock{{\it J. Fluid. Mech. } {\bf 3} pp.~494--514.}

\bibitem{Batch57} Batchelor, G.~K.:1957,
\newblock {`Wave Scattering Due to Turbulence',}
\newblock {in {\it Symposium on Naval Hydrodynamics} F.~S. Sherman (ed.), National Academy of Sciences, Washington pp.~403--429.}

\bibitem{Lund89} Lund, F. and C. Rojas: 1989,
\newblock {`Ultrasound as a probe of turbulence',}
\newblock {{\it Physica D} {\bf 37}, pp.~508--514.}

\bibitem{Llewellyn1} Llewellyn~Smith, S.~G., and R. Ford,: 2001,
\newblock {`Three-dimensional acoustic scattering by vortical flows. I. General theory',}
\newblock {{\it Phys. Fluids} {\bf 13}, 10, pp.~2876--2889.}

\bibitem{Llewellyn2} Llewellyn~Smith, S.~G., and R. Ford,: 2001,
\newblock {`Three-dimensional acoustic scattering by vortical flows. II. Axisymmetric scattering by Hill's spherical vortex',}
\newblock {{\it Phys. Fluids} {\bf 13}, 10, pp.~2890--2900.}

\bibitem{Colonius94} Colonius, T., S.ÊK. Lele, P. Moin: 1994,
\newblock {`The scattering of sound waves by a vortex: numerical and analytical solutions',}
\newblock {{\it J. Fluid Mech.} {\bf 260}, 10, pp.~271--.}

\bibitem{Papoulis} Papoulis, A.:
\newblock {{\em Signal Analysis}}, 
\newblock {McGraw-Hill, 1984.}

\bibitem{Kinsler82} Kinsler, L.~E., A.~R. Frey, A.~B. Coppens, and J.~V. Sanders: 
\newblock {{\em Fundamentals of Acoustics}}, 
\newblock {Wiley and Sons, Fourth edition, 2000.}

\bibitem{Malecot} Malecot, Y., C.~Auriault, H.~Kahalerras, Y.~Gagne, O.~Chanal,~B. Chabaud, and B.~Castaing: 2000,
\newblock {`A statistical estimator of turbulence intermittency in physical and numerical experiments.'}.
\newblock {{\it Eur. Phys. J. B}, {\bf 16}, pp.~549--561.}

\bibitem{GReC_Adv_Cryo_2002}
Bezaguet A., J.-P. Dauvergne, S. Knoops, P. Lebrun ...  C. Baudet, Y.Gagne, C. Poulain, B. Castaing, Y. Ladam and F. Vittoz: 2002,
\newblock {`A cryogenic high Reynolds turbulence experiment at CERN'}.
\newblock {{\it Adv. Cryo. Eng.}, {\bf 47}, pp.~136-144.}

\bibitem{GReC_PhysicaC_2003}
Pietropinto S., C. Poulain, C. Baudet, B. Castaing, B. Chabaud, Y. Gagne, B. Hebral, Y. Ladam, P. Lebrun, 
O. Pirotte and P. Roche: 1999,
\newblock {`Superconducting instrumentation for high Reynolds turbulence experiments with low temperature gaseous helium.'}.
\newblock {{\it Physica C}, {\bf 386}, pp.~512--516.}

\bibitem{Anke74} D.Anke, Acustica {\bf 30}, (1974).

\bibitem{Bendat_Piersol} Bendat, J.~S., .~G. Piersol:
\newblock {{\em Random Data. Analysis and mesurement procedures}}, 
\newblock {John Wiley and Sons, 1986.}

\bibitem{Wallace86} Wallace, J.~M.: 1986, 
\newblock {`Methods for measuring vorticity in turbulent flows',}
\newblock {{\it Experiments in Fluids} {\bf 4}, pp.~61--71.}

\bibitem{Tsin92} Tsinober, A., E. Kit, and T. Dracos: 1992, 
\newblock {`Experimental investigation of the field of velocity gradients in turbulent flows',} 
\newblock {{\it J. Fluid. Mech.}, {\bf 242}, pp.~169--192.}

\bibitem{Anton97} Shafi, H.~S., and R.~A. Antonia: 1997, 
\newblock {`Small-scale characteristics of a turbulent boundary layer over a rough wall',}
\newblock {{\it J. Fluid. Mech.}, {\bf 342}, pp.~263--293.}

\bibitem{Tennekes} Tennekes, H., and J.~L. Lumley:
\newblock {{\em A First Course in Turbulence}}, 
\newblock {The MIT Press, 1972.}

\bibitem{Batch53} Batchelor G.K.:
\newblock {{\em The Theory of Homogeneous Turbulence}},
\newblock {Cambridge University Press, 1953.}

\bibitem{Pumir2001} Pumir, A.: 2001,
\newblock {Alain Pumir, (private communication).}

\bibitem{Flandrin99} Flandrin, P.:
\newblock {{\em Time-Frequency/Time-Scale Analysis}}
\newblock {Academic Press, 1999.}

\bibitem{Will92} Williams W.~J., J.~Jeong: 1992,
\newblock {in {\em Time-frequency signal analysis: methods and applications}}
\newblock {B.~Bouashash Ed, Longman \& Cheshire, Chapter~3, pp.~74--97.}

\bibitem{Baudet99}
Baudet, C., O. Michel, and W.~J. Williams.: 1999,
\newblock {`Detection of Coherent Vorticity Structures using Time-Scale Resolved Acoustic Spectroscopy'}.
\newblock {{\it Physica D}, {\bf 128}, pp.~1--17.}

\bibitem{Sanada92} Sanada, T., and V. Shanmugasundaram: 1992,
\newblock {`Random sweeping effect in isotropic numerical turbulence',}
\newblock {{\it Phys. Fluids} {\bf 4} (6), pp.~1245--1250.}

\bibitem{Xu01} Xu G., R.~A. Antonia, and S. Rajagopalan: 2001,
\newblock {`Sweeping decorrelation hypothesis in a turbulent round jet.'}.
\newblock {{\it  Fluid. Dyn. Res.} {\bf 28}, pp.~311--321}

\bibitem{Novikov93} Novikov E.: 1993,
\newblock {`Vortical scales for two-and three-dimensional turbulence',}
\newblock {{\it Phys. Rev. E} {\bf 49} (2), pp.~975--977.}

\bibitem{Novikov94} Novikov E.: 1994,
\newblock {`Statistical balance of Vorticity and a new scale for vortical structures in turbulence',}
\newblock {{\it Phys. Rev. Lett.} {\bf 71} (17), pp~2718--20.}

\bibitem{Pinton01} Mordant N., Metz P., Michel O., and Pinton J.-F.: 2001, 
\newblock {`Measurements of Lagranfian velocity in fully developed turbulence'}
\newblock {{\it Phys. Rev. Lett.} {\bf 87} (25), 214501.}

\bibitem{Pinton02} Mordant N., Pinton J.-F., ans Michel O.,: 2002, 
\newblock {`Time resolved tracking of a sound scatterer in a complex flow: non-stationary signal analysis and applications'}
\newblock{{\it J.Acoust. Soc. Am. } {\bf 112} (1) pp.~108--1118.}
  
\bibitem{Voth98} Voth G.~A., K. Satyanarayan, and E. Bodenschatz: 1998, 
\newblock {`Lagrangian acceleration measurements at large Reynolds numbers'}
\newblock {{\it Phys. Fluids} {\bf 10} (9) pp.~2268--2280.}

\bibitem{Pope2000} Pope, S.~B.: 
\newblock {{\em Turbulent flows}},
\newblock {Cambridge University Press, 2000.}

\bibitem{Baudet91} Baudet, C., S. Ciliberto and J-F. Pinton: 1991,
\newblock {`Spectral analysis of the von K\'arm\'an flow using ultrasound scattering',}
\newblock {{\it Phys.Rev.Lett.}, {\bf 67} (2), pp.~193--195.}

\bibitem{Gromov82} Gromov, P.~R., A.~B. Ezerskii and A.~L. Fabrikant: 1982, 
\newblock {`Sound scattering by a vortex wake behind a cylinder',}
\newblock {{\it Sov.Phys.Acoust.}, {\bf 28-6}, pp.~452--455.}

\bibitem{Dubrulle03} Dubrulle B., J-P. Laval, S. Nazarenko, and O. Zaboronski: 2003,
\newblock {`A model for rapid stochastic distorsions of small-scale turbulence',}
\newblock {{\it Submitted to J. Fluid Mech.}, http://arXiv:physics/0304035}

\bibitem{Nazarenko03} Nazarenko, S., R.~J. West and O. Zaboronski: 2003,
\newblock {`Statistics of fourier modes in the Kazantsev-Kraichnan dynamo model',}
\newblock {{\it Submitted to Phys. Rev. E}}

\end{thebibliography}
\end{document}